# Realization of a $\mathbb{Z}$-classified chiral-symmetric higher-order topological insulator in a coupling-inverted acoustic crystal

Dongyi Wang*[1], Yuanchen Deng*[2], Jun Ji[2], Mourad Oudich[2], Wladimir A. Benalcazar[†3], Guancong Ma[†1], Yun Jing[†2]

[1]Department of Physics, Hong Kong Baptist University, Kowloon Tong, Hong Kong, China
[2]Graduate Program in Acoustics, The Pennsylvania State University, University Park, Pennsylvania 16802, USA
[3]Department of Physics, Emory University, Atlanta, GA 30322, USA

*D.W. and Y.D. contributed equally to this work.
[†]Email: yqj5201@psu.edu, phgcma@hkbu.edu.hk, benalcazar@emory.edu

**Abstract**

Higher-order topological band theory has transformed the landscape of topological phases in quantum and classical systems. Here, we experimentally demonstrate a two-dimensional higher-order topological phase, referred to as the multiple chiral topological phase, which is protected by a multipole chiral number (MCN). Our realization differs from previous higher-order topological phases in that it possesses a larger-than-unity MCN, which arises when the nearest-neighbor couplings are weaker than long-range couplings. Our phase has an MCN of 4, protecting the existence of 4 midgap topological corner modes at each corner. The multiple topological corner modes demonstrated here could lead to enhanced quantum-inspired devices for sensing and computing. Our study also highlights the rich and untapped potential of long-range coupling manipulation for future research in topological phases.

Recent theoretical advancements in higher-order topological phases (HOTPs) [1–8] have substantially expanded the scope of the bulk-boundary correspondence, leading to the observation of topological corner modes (TCMs) [9–18] and topological defect modes [19–23]. TCMs are zero-dimensional bound states localized at the corners of a HOTP specimen, which have inspired novel applications such as topological lasers [24], thermal engineering [25], and quantum optics [26]. In 2D, one can obtain HOTPs by either engineering the Wannier center configuration of the lattice or the existence of boundary-localized mass domains. Both approaches give rise to only one TCM at each corner. Yet, the presence of multiple TCMs at a single corner confers several advantages, notably including the possibility of realizing multimode topological lasers. Furthermore, the existence of multiple degenerate or nearly degenerate TCMs holds the potential for investigating the dynamics of non-Abelian characteristics [27,28].

As a result, the pursuit of an efficient method to create the desired quantity of in-gap TCMs remains a highly coveted objective.

The band topology of a standard one-dimensional Su-Schrieffer-Heeger model can be characterized by a winding number that is either zero or one [29]. However, by the introduction of proper long-range couplings, the winding number can be increased. Such systems, therefore, belong to a $\mathbb{Z}$-classified topological phase (See Ref. [30] for more details). In fact, topological phases that protect multiple states at each 0D boundary exist in odd-dimensional chiral-symmetric systems, protected by the winding number of their Bloch Hamiltonians across the Brillouin zone. A recent theoretical study found a $\mathbb{Z}$-classification of 2 HOTPs in class AIII by an unconventional generalization of the "winding number" to higher dimensions [31,35–37]. The topological invariants are referred to as the "multipole chiral numbers" (MCN), as they are built from sublattice multipole moment operators. A lattice with MCNs greater than one can safeguard multiple zero-energy states per corner, all of which are pinned at midgap (zero energy) due to chiral symmetry. This phenomenon manifests itself in a tight-binding model wherein the long-range couplings (LRCs) are greater than nearest-neighbor couplings (NNCs) in magnitude. This condition, which we call "coupling inversion," is hard to obtain in natural materials because it goes against the general decay rule of bound electronic wavefunctions. Thus, the experimental realization of chiral-symmetric HOTPs with MCN>1 is unlikely in condensed matter systems.

We thus turn our attention to classical-wave systems, which have become versatile testbeds for topological models [38–40]. For example, acoustic crystals built from coupled acoustic cavities offer a powerful platform for studying HOTPs with advanced designs [10,11,16,17,22,41]. In addition, it has been systematically established that chiral symmetry can be precisely emulated in such systems [41], which makes it possible to realize topological modes pinned at zero energy.

In this letter, we leverage the salient features of coupled acoustic cavities to experimentally realize a chiral-symmetric acoustic MCTP with an MCN of 4. By using judiciously designed space-coiling channels, we show that NNC and LRC coefficients are precisely implemented while preserving chiral symmetry. More importantly, the lattice features coupling inversion by having LRCs stronger than NNCs. Our work demonstrates experimentally the existence of a chiral-symmetric HOTP with MCN>1 for the first time and highlights the important role of LRCs in realizing novel topological phases; it also exemplifies the great potential of acoustic platforms for studying novel topological models.

The system of interest is a square lattice of 4-site unit cells, described by the tight binding Hamiltonian

$$H(\mathbf{k}) = \begin{bmatrix} 0 & h(\mathbf{k}) + g(\mathbf{k}) \\ h^\dagger(\mathbf{k}) + g^\dagger(\mathbf{k}) & 0 \end{bmatrix}, \qquad (1)$$

where

$$h(\mathbf{k}) = \begin{pmatrix} u + ve^{-ik_x} & -u - ve^{-ik_y} \\ -u - ve^{ik_y} & -u - ve^{ik_x} \end{pmatrix}, \; g(\mathbf{k}) = \begin{pmatrix} we^{-2ik_x} & -we^{-2ik_y} \\ -we^{2ik_y} & -we^{2ik_x} \end{pmatrix}, \qquad (2)$$

with $u, v$, being the magnitudes of the intracell and intercell nearest-neighbor couplings, respectively, and $w$ being the magnitude of the LRCs in $x$ and $y$ directions. Omitting the matrix $g(k)$ momentarily, the Hamiltonian describes a quadrupole topological insulator (QTI) when $v > u$, which is characterized by a non-trivial quadrupole moment $q_{xy} = \frac{1}{2}$ [1,3]. Due to its chiral symmetry, it also possesses an MCN. In 2D, the MCN is defined as $N_{xy} = \frac{1}{2\pi i}\text{Tr}\big[\log\big(\bar{Q}_{xy}^A \bar{Q}_{xy}^{B\dagger}\big)\big]$, where $\bar{Q}_{xy}^A$ and $\bar{Q}_{xy}^{B\dagger}$ are multipole moment operators for sublattices A and B [31]. The MCN is $N_{xy} = 1$ in this case, resulting in one zero-energy topological mid-gap state per corner of a finite-size rectangular lattice.

Reintroducing $g(k)$, which contains only LRCs, the lattice takes the form shown in Fig. 1 (a). The LRCs are arranged in a way that an additional synthetic magnetic field with $\pi$-flux threads each plaquette in the lattice. The system becomes an MCTP with an MCN of $N_{xy} = 4$ when coupling inversion occurs, i.e., when $w > v$ (see Supplemental Material [30] for further analysis). Figure 1(b) shows the spectrum of a $16 \times 16$-site square lattice. A bulk gap is opened near the zero energy, wherein sixteen near-degenerate in-gap modes are observed. These are bound states localized around the four corners (four per corner), as shown in Fig. 1(c). These are the TCMs protected by the non-zero MCN. Because of the presence of the LRCs, some TCMs are localized several sites away from the extreme corner. We note that the energies of the TCMs are not exactly degenerate but remain symmetric about zero energy, which is due to the finite-size effect. The TCMs are robust against reasonable disorders (see Ref. [30] for details), providing tolerance for fabrication and experiment errors. Finally, the phase diagram of the MCTP is plotted and can be found in [30].

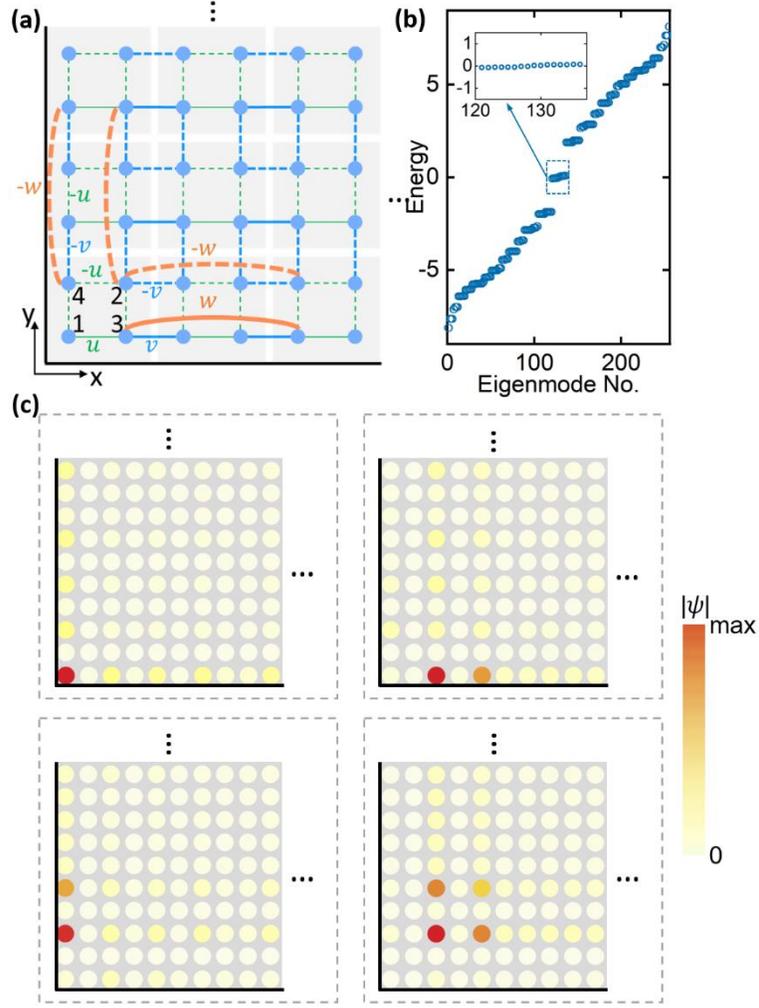

FIG. 1 A schematic and the corresponding theoretical results regarding the topological corner modes (TCMs) obtained from the tight-binding model. (a) The schematic shows the lattice model and the configuration near a corner. The solid (dashed) lines represent positive (negative) couplings, and their thickness indicates the coupling strength. Colors are also used to denote different couplings, including intracell NNCs (green), intercell NNCs (blue), and LRCs (orange). Not all the LRCs are shown for the sake of clarity. (b) Sixteen mid-gap TCMs. (c) The wavefunction amplitudes of the four TCMs in one of the four corners (bottom left corner) of the lattice. Here, $u, v, w = 1,2,3$.

Next, we design an acoustic crystal to realize the model [Eq. (1)]. Our starting point is a 2D array of coupled acoustic cavities [31]. The cavities are identical air-filled cuboids with a square cross-section (in the *xy* plane) and a relatively large height (along the *z* axis). The cavity mode of interest is the first-order resonance mode, whose natural frequency is $f_0 = \frac{c_0}{2L_0} = 2858$ Hz, and corresponds to a modal profile $P(x, y, z) = P_0 \cos\left(\frac{2\pi z}{L_0}\right)$, where $c_0 = 343\ m/s$ is the speed of sound in air, $P_0$ is the pressure amplitude, and $L_0 = 60\ mm$ is the cavity height. The coupling among cavities is facilitated using air channels, which are acoustic waveguides with subwavelength cross-sectional areas. A two-cavity setup is used to identify the conditions needed to build the model [Fig. 2(a)], which is described by a two-state Hamiltonian $H_{2s} = \begin{pmatrix} 2\pi f_0 & t \\ t & 2\pi f_0 \end{pmatrix}$, where $t \in \mathbb{R}$ is the coupling term. This Hamiltonian possesses an even mode and an odd mode, with eigenfrequencies $2\pi f_e$ and $2\pi f_o$, respectively. Their normalized splitting, defined as $\Delta f = \frac{f_e - f_o}{f_0}$, indicates the coupling strength and coupling sign. From the parity of the modes, it is straightforward that $\text{sgn}(t) = \text{sgn}(\Delta f)$. On the other hand, when the channel produces no perturbation to the natural frequency of the cavities, chiral symmetry is respected, and $f_e$ and $f_o$ are symmetric about $f_0$. This $f_0$ is regarded as the "zero energy." We use the normalized eigenvalue mean, $\bar{f} = \frac{f_e + f_o}{2f_0}$, to quantify the breaking of chiral symmetry. Chiral symmetry is strictly respected when $\bar{f} = 1$. We compute $\Delta f$ and $\bar{f}$ using COMSOL Multiphysics. The tuned parameters are channel length $L_c$, width $D_c$, and connecting position $z_c$ on the cavity. The results are shown in Fig. 2(b) and 2(c). According to the configuration of the unit cell [Fig. 1], we need six different coupling coefficients, four for NNCs, and two for LRCs (see Supplemental Material [30] for details of coupling channel design). The chosen parameters are correspondingly indicated by colored markers (diamonds and stars) in Fig. 2(b) and 2(c).

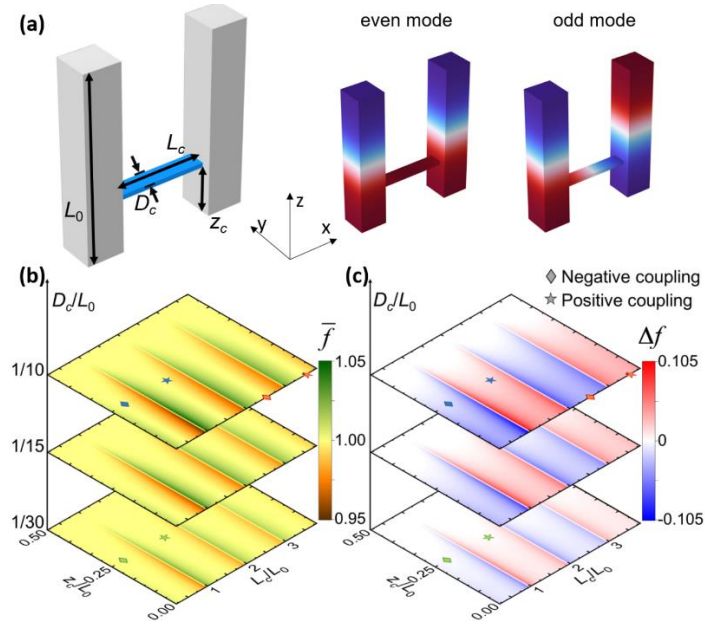

FIG. 2. (a) The eigenmodes of the two acoustic cavities coupled with an air channel. The normalized eigenvalue mean $\bar{f}$ (b) and eigenvalue splitting $\Delta f$ (c) as functions of $L_c$ and $z_c$, and for different $D_c$. The diamonds and stars mark the geometric values chosen for our acoustic HOTP design.

Using these geometric parameters, the acoustic crystal design is shown in Fig. 3(a). The key requirement, coupling inversion, is achieved by using LRC channels connected at the top and bottom of the cavities where the pressure amplitude (and therefore the coupling) is maximum, while their NNC counterparts are connected at ¼ height of the cavities (leading to weaker coupling). The lengths of the LRC channels are $\sim 2.5 L_0$ and $\sim 3.5 L_0$ for negative and positive coupling, respectively, which are longer than the spatial separation of the relevant cavities. The coupling channels are essentially 1D acoustic waveguides with subwavelength cross-sectional dimensions. Space-coiling design [34,42], which is widely used in acoustic metamaterials and metasurfaces [43,44], is employed to bend the waveguides into meandering shapes so that the coupling channels can fit the cavity array. The detailed geometric parameters are presented in [30]. We then perform finite-element simulations on the design, and the results are shown in Fig. 3(b) and 3(c), which can be directly compared with the theoretical results shown in Fig. 1(b) and 1(c). Sixteen mid-gap TCMs can be identified, all of which localize at corners, which is compatible with a crystal in the MCN=4 phase. In addition, both the bulk bands and the TCMs are nearly symmetric about 2858 Hz, and the TCMs are safely separated from the bulk bands. The splitting of the TCMs away from mid-gap is attributed to hybridization due to mode overlap across opposite corners and intrinsic deviations away from perfect chiral symmetry. For the practical purposes of topological protection, all the TCMs are robustly protected by approximate chiral symmetry. Furthermore, since the hybridization due to mode overlap is a finite-size effect, it will decrease with increasing system size.

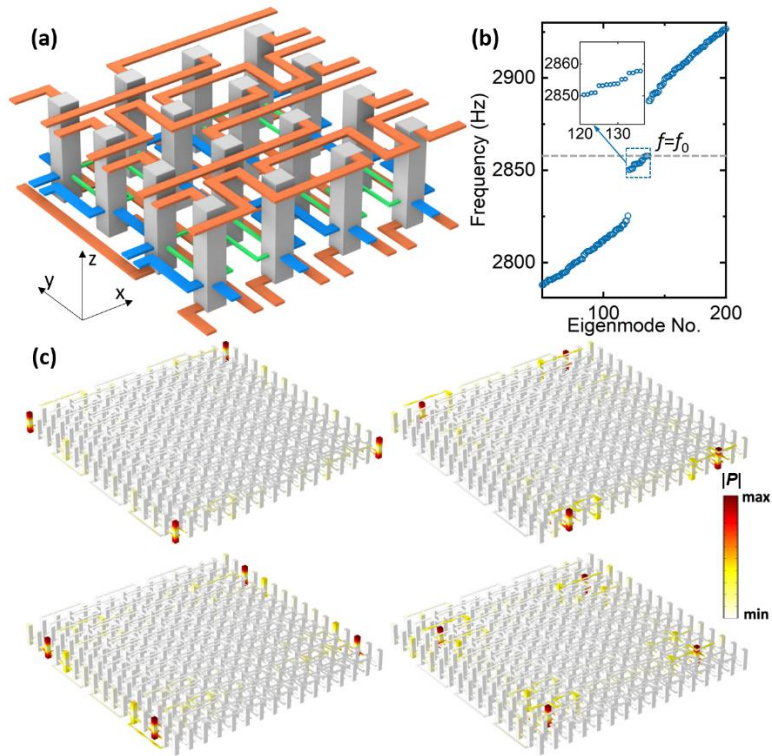

FIG. 3. (a) The schematic drawing of the acoustic lattice design. The colors of the air channels are in one-to-one correspondence with those in Fig. 1. (b) The simulated eigenfrequencies of the acoustic lattice. (c) Sixteen mid-gap TCMs are identified. They are separated into four groups based on their spatial profiles.

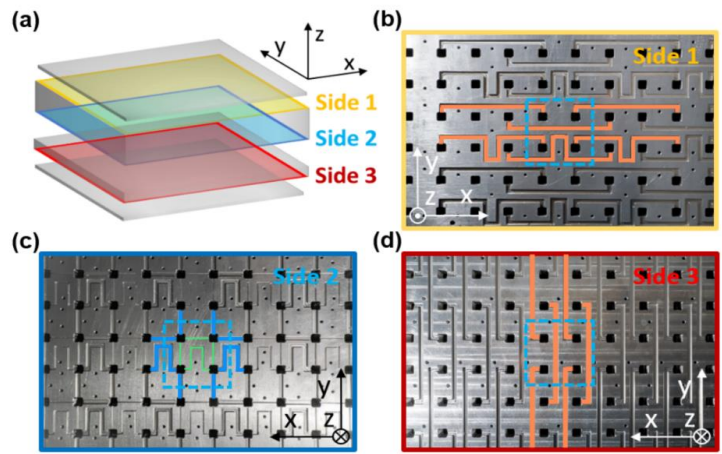

FIG. 4. The fabricated acoustic lattice used in our experiments. (a) A schematic showing the design. Different sets of air channels are machined on three different sides, as shown in (b, c, d), respectively.

The fabricated acoustic crystal is shown in Fig. 4. It is composed of two aluminum blocks and two cover plates [Fig. 4(a)]. The lattice has $16 \times 16$ cavities, and the air channels are etched on different surfaces of the aluminum blocks using CNC machining [Fig. 4(b)-4(d)]. Holes are opened on the top cover plates for excitation and measurement. They are sealed with silastic plugs when not in use. For the acoustic measurement, a short pulse covering 2200–3500 Hz is sent using a 16-channel sound card (MOTU 16A) through a power amplifier to drive a loudspeaker placed at the positions illustrated in Fig. 5(b) for the excitation of the bulk mode and TCMs. The response signals are detected by 16 identical microphones and recorded by the sound card. The results are shown in Fig. 5(a) and 5(b). A bulk gap is clearly observed in the frequency range of 2800–2900 Hz. When the loudspeakers are placed at the designated positions around the corners [indicated by the red arrows in Fig. 5(b)], 16 response peaks are observed in the bulk gap (near centre frequency $f_0 = 2858$ Hz) when excited at different positions marked 1~16 as shown in Fig. 5(b). We note that the responses peak at slightly different frequencies. This is attributed to fabrication and experiment error, yet the MCTP is robust against such disorders as mentioned in previous sections. The spatial profiles of these responses are mapped out across the entire lattice, as shown in Fig. 5(b). They are in good agreement with the numerically obtained eigenfunctions of TCMs [Fig. 3(c)]. The slight deviations in peak frequencies of the TCM responses are attributed to fabrication errors. The maximum deviation is $\pm 16$ Hz, which is about 0.56% for a base frequency of 2858 Hz. Overall, the successful realization of MCTP is collectively demonstrated by the observation of a total of 16 TCMs in the experiment, and the fact that, not only the number of the TCMs, but also their eigenfunctions are in excellent agreement with the predictions made by numerical simulations based on the same acoustic crystal. Discussion and results on the effect of thermoviscous losses are included in [30].

In conclusion, we have experimentally demonstrated a coupling-inverted acoustic crystal possessing MCTP that is characterized by a $\mathbb{Z}$ topological invariant, the MCN, which is greater than 1. Unlike most topological models, the existence of MCTP explicitly requires the presence of LRCs. Our work provides strong evidence that the manipulation of LRCs can be a fruitful route for accessing new topological phases. Because of the difficulty in finding LRC in natural materials, classical-wave platforms such as acoustic crystals may play an increasingly important role in future studies of topological phases, particularly in those that additionally require chiral symmetry. Meanwhile, it would be interesting to expand MCTP into other systems, such as photonic crystals and topoelectrical circuits, though it remains unclear whether chiral symmetry can be achieved in photonic crystals (periodic arrays of coupled waveguides do possess chiral symmetry [14]). Having spatial and spectral coexistence of multiple TCMs, our MCTP may be beneficial for applications such as spatially multiplexed energy confinement [45], multi-mode corner-emitting topological lasers [24], and may open a route for the non-Abelian manipulation of TCMs [46–48].

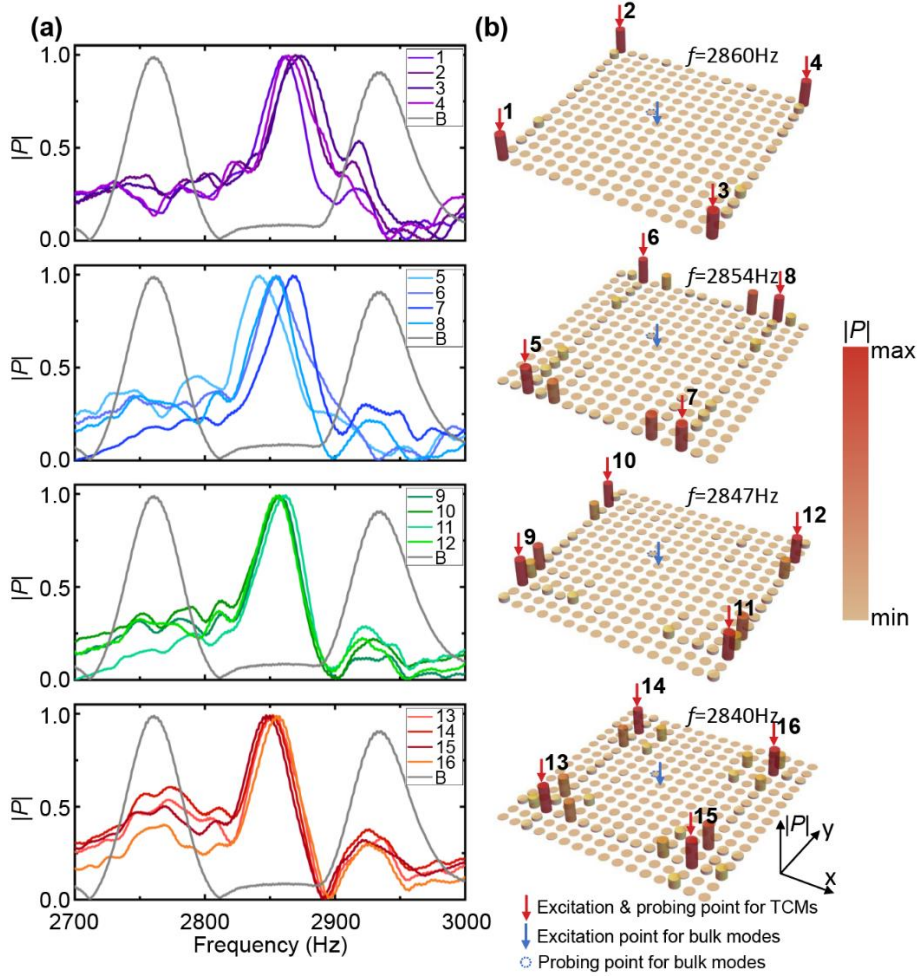

FIG. 5. Experimental results. (a) Response spectra of the bulk modes (gray curves, denoted B in the legends) and TCMs (colored curves) marked with numbers 1-16, corresponding to different excitation positions shown by the arrows in (b). Responses of the TCMs are presented in four subfigures on the right, each representing a specific category of TCMs determined by their positions relative to the exact corner site (e.g., exactly at the corner, two sites away from the corner on the x-axis, etc.). (b) The spatial response profiles with excitation and measurement positions at the corners (marked by the red arrows). The blue arrows and circles (right next to the blue arrows) mark the excitation and probing points for the bulk spectra in (a).

*Acknowledgments.* Y. J. thanks the NSF for support through CMMI-2039463. W. A. B. acknowledges the support from the startup fund at Emory University. G. M. is supported by the Hong Kong Research Grants Council (RFS2223-2S01, 12301822, 12302420) and the National Key R&D Program of China (2022YFA1404400). D. W. thanks Wei Wang and Tong Liu for discussions on theoretical parts, and thanks Xulong Wang for assisting with the experiment.

**References**
[1] W. A. Benalcazar, B. A. Bernevig, and T. L. Hughes, *Quantized Electric Multipole Insulators*, Science **357**, 61 (2017).
[2] Z. Song, Z. Fang, and C. Fang, *( D − 2 ) -Dimensional Edge States of Rotation Symmetry Protected Topological States*, Phys. Rev. Lett. **119**, 246402 (2017).
[3] W. A. Benalcazar, B. A. Bernevig, and T. L. Hughes, *Electric Multipole Moments, Topological Multipole Moment Pumping, and Chiral Hinge States in Crystalline Insulators*, Phys. Rev. B **96**, 245115 (2017).
[4] F. Schindler, A. M. Cook, M. G. Vergniory, Z. Wang, S. S. P. Parkin, B. A. Bernevig, and T. Neupert, *Higher-Order Topological Insulators*, SCIENCE ADVANCES (2018).
[5] C. W. Peterson, T. Li, W. A. Benalcazar, T. L. Hughes, and G. Bahl, *A Fractional Corner Anomaly Reveals Higher-Order Topology*, Science **368**, 1114 (2020).
[6] B. Xie, H.-X. Wang, X. Zhang, P. Zhan, J.-H. Jiang, M. Lu, and Y. Chen, *Higher-Order Band Topology*, Nat Rev Phys **3**, 520 (2021).
[7] T. Li, P. Zhu, W. A. Benalcazar, and T. L. Hughes, *Fractional Disclination Charge in Two-Dimensional C n -Symmetric Topological Crystalline Insulators*, Phys. Rev. B **101**, 115115 (2020).
[8] Y. Ran, Y. Zhang, and A. Vishwanath, *One-Dimensional Topologically Protected Modes in Topological Insulators with Lattice Dislocations*, Nature Phys **5**, 298 (2009).
[9] M. Serra-Garcia, V. Peri, R. Süsstrunk, O. R. Bilal, T. Larsen, L. G. Villanueva, and S. D. Huber, *Observation of a Phononic Quadrupole Topological Insulator*, Nature **555**, 342 (2018).
[10] H. Xue, Y. Yang, F. Gao, Y. Chong, and B. Zhang, *Acoustic Higher-Order Topological Insulator on a Kagome Lattice*, Nature Mater **18**, 108 (2019).
[11] X. Ni, M. Weiner, A. Alù, and A. B. Khanikaev, *Observation of Higher-Order Topological Acoustic States Protected by Generalized Chiral Symmetry*, Nature Mater **18**, 113 (2019).
[12] C. W. Peterson, W. A. Benalcazar, T. L. Hughes, and G. Bahl, *A Quantized Microwave Quadrupole Insulator with Topologically Protected Corner States*, Nature **555**, 346 (2018).
[13] S. Imhof et al., *Topolectrical-Circuit Realization of Topological Corner Modes*, Nature Phys **14**, 925 (2018).
[14] J. Noh, W. A. Benalcazar, S. Huang, M. J. Collins, K. P. Chen, T. L. Hughes, and M. C. Rechtsman, *Topological Protection of Photonic Mid-Gap Defect Modes*, Nature Photon **12**, 408 (2018).
[15] X. Zhang, H.-X. Wang, Z.-K. Lin, Y. Tian, B. Xie, M.-H. Lu, Y.-F. Chen, and J.-H. Jiang, *Second-Order Topology and Multidimensional Topological Transitions in Sonic Crystals*, Nat. Phys. **15**, 582 (2019).
[16] Y. Qi, C. Qiu, M. Xiao, H. He, M. Ke, and Z. Liu, *Acoustic Realization of Quadrupole Topological Insulators*, Phys. Rev. Lett. **124**, 206601 (2020).
[17] Z.-G. Chen, W. Zhu, Y. Tan, L. Wang, and G. Ma, *Acoustic Realization of a Four-Dimensional Higher-Order Chern Insulator and Boundary-Modes Engineering*, Phys. Rev. X **11**, 011016 (2021).
[18] W. Wang, Z.-G. Chen, and G. Ma, *Synthetic Three-Dimensional Z × Z 2 Topological Insulator in an Elastic Metacrystal*, Phys. Rev. Lett. **127**, 214302 (2021).
[19] Y. Liu, S. Leung, F.-F. Li, Z.-K. Lin, X. Tao, Y. Poo, and J.-H. Jiang, *Bulk–Disclination Correspondence in Topological Crystalline Insulators*, Nature **589**, 381 (2021).
[20] C. W. Peterson, T. Li, W. Jiang, T. L. Hughes, and G. Bahl, *Trapped Fractional Charges at Bulk Defects in Topological Insulators*, Nature **589**, 376 (2021).
[21] Y. Deng, W. A. Benalcazar, Z.-G. Chen, M. Oudich, G. Ma, and Y. Jing, *Observation of Degenerate Zero-Energy Topological States at Disclinations in an Acoustic Lattice*, Phys. Rev. Lett. **128**, 174301 (2022).
[22] Z.-K. Lin, Y. Wu, B. Jiang, Y. Liu, S.-Q. Wu, F. Li, and J.-H. Jiang, *Topological Wannier Cycles Induced by Sub-Unit-Cell Artificial Gauge Flux in a Sonic Crystal*, Nat. Mater. **21**, 430 (2022).
[23] L. Ye, C. Qiu, M. Xiao, T. Li, J. Du, M. Ke, and Z. Liu, *Topological Dislocation Modes in Three-Dimensional Acoustic Topological Insulators*, Nat Commun **13**, 508 (2022).

[24] W. Zhang et al., *Low-Threshold Topological Nanolasers Based on the Second-Order Corner State*, Light Sci Appl **9**, 109 (2020).
[25] G. Xu, X. Zhou, S. Yang, J. Wu, and C.-W. Qiu, *Observation of Bulk Quadrupole in Topological Heat Transport*, Nat Commun **14**, 3252 (2023).
[26] C. Li, M. Li, L. Yan, S. Ye, X. Hu, Q. Gong, and Y. Li, *Higher-Order Topological Biphoton Corner States in Two-Dimensional Photonic Lattices*, Phys. Rev. Research **4**, 023049 (2022).
[27] P. Boross, J. K. Asbóth, G. Széchenyi, L. Oroszlány, and A. Pályi, *Poor Man's Topological Quantum Gate Based on the Su-Schrieffer-Heeger Model*, Phys. Rev. B **100**, 045414 (2019).
[28] Y. Wu, H. Jiang, J. Liu, H. Liu, and X. C. Xie, *Non-Abelian Braiding of Dirac Fermionic Modes Using Topological Corner States in Higher-Order Topological Insulator*, Phys. Rev. Lett. **125**, 036801 (2020).
[29] J. K. Asbóth, L. Oroszlány, and A. Pályi, *A Short Course on Topological Insulators*, Vol. 919 (Springer International Publishing, Cham, 2016).
[30] *See Supplemental Material at Http://Link.Aps.Org/Supplemental/10.1103/PhysRevLett.131.157201 for Phase Diagram and Further Analysis of the MCTP, Design of the Coupling Channels, and Evaluations on Robustness against Disorder and Effect of Loss, Which Includes Refs. [16,31–34]*.
[31] W. A. Benalcazar and A. Cerjan, *Chiral-Symmetric Higher-Order Topological Phases of Matter*, Phys. Rev. Lett. **128**, 127601 (2022).
[32] W. Zhu and G. Ma, *Distinguishing Topological Corner Modes in Higher-Order Topological Insulators of Finite Size*, Phys. Rev. B **101**, 161301 (2020).
[33] X. Li, S. Wu, G. Zhang, W. Cai, J. Ng, and G. Ma, *Measurement of Corner-Mode Coupling in Acoustic Higher-Order Topological Insulators*, Front. Phys. **9**, 770589 (2021).
[34] Z. Liang and J. Li, *Extreme Acoustic Metamaterial by Coiling Up Space*, Phys. Rev. Lett. **108**, 114301 (2012).
[35] A. Kitaev, V. Lebedev, and M. Feigel'man, *Periodic Table for Topological Insulators and Superconductors*, in *AIP Conference Proceedings* (AIP, Chernogolokova (Russia), 2009), pp. 22–30.
[36] S. Ryu, A. P. Schnyder, A. Furusaki, and A. W. W. Ludwig, *Topological Insulators and Superconductors: Tenfold Way and Dimensional Hierarchy*, New J. Phys. **12**, 065010 (2010).
[37] C.-K. Chiu, J. C. Y. Teo, A. P. Schnyder, and S. Ryu, *Classification of Topological Quantum Matter with Symmetries*, Rev. Mod. Phys. **88**, 035005 (2016).
[38] T. Ozawa et al., *Topological Photonics*, Rev. Mod. Phys. **91**, 015006 (2019).
[39] G. Ma, M. Xiao, and C. T. Chan, *Topological Phases in Acoustic and Mechanical Systems*, Nat Rev Phys **1**, 281 (2019).
[40] H. Xue, Y. Yang, and B. Zhang, *Topological Acoustics*, Nat Rev Mater **7**, 974 (2022).
[41] Z.-G. Chen, L. Wang, G. Zhang, and G. Ma, *Chiral Symmetry Breaking of Tight-Binding Models in Coupled Acoustic-Cavity Systems*, Phys. Rev. Applied **14**, 024023 (2020).
[42] Y. Xie, B.-I. Popa, L. Zigoneanu, and S. A. Cummer, *Measurement of a Broadband Negative Index with Space-Coiling Acoustic Metamaterials*, Phys. Rev. Lett. **110**, 175501 (2013).
[43] Y. Xie, W. Wang, H. Chen, A. Konneker, B.-I. Popa, and S. A. Cummer, *Wavefront Modulation and Subwavelength Diffractive Acoustics with an Acoustic Metasurface*, Nat Commun **5**, 5553 (2014).
[44] Y. Li, X. Jiang, R. Li, B. Liang, X. Zou, L. Yin, and J. Cheng, *Experimental Realization of Full Control of Reflected Waves with Subwavelength Acoustic Metasurfaces*, Phys. Rev. Applied **2**, 064002 (2014).
[45] Y. Ota, F. Liu, R. Katsumi, K. Watanabe, K. Wakabayashi, Y. Arakawa, and S. Iwamoto, *Photonic Crystal Nanocavity Based on a Topological Corner State*, Optica **6**, 786 (2019).
[46] Z.-G. Chen, R.-Y. Zhang, C. T. Chan, and G. Ma, *Classical Non-Abelian Braiding of Acoustic Modes*, Nat. Phys. **18**, 179 (2022).
[47] X.-L. Zhang, F. Yu, Z.-G. Chen, Z.-N. Tian, Q.-D. Chen, H.-B. Sun, and G. Ma, *Non-Abelian Braiding on Photonic Chips*, Nat. Photon. **16**, 390 (2022).

[48] Y.-K. Sun, X.-L. Zhang, F. Yu, Z.-N. Tian, Q.-D. Chen, and H.-B. Sun, *Non-Abelian Thouless Pumping in Photonic Waveguides*, Nat. Phys. **18**, 1080 (2022).